\documentclass[%
 aip,
 amsmath,amssymb,
 reprint,%
]{revtex4-1}

\usepackage{graphicx}
\usepackage{placeins}
\usepackage{siunitx}
\renewcommand{\topfraction}{0.9}
\renewcommand{\dbltopfraction}{0.9}
\renewcommand{\textfraction}{0.07}
\renewcommand{\floatpagefraction}{0.75}
\renewcommand{\dblfloatpagefraction}{0.75}
\graphicspath{{Figures/}{./}}
\usepackage{dcolumn}
\usepackage{bm}

\usepackage[utf8]{inputenc}
\usepackage[T1]{fontenc}
\usepackage{mathptmx}
\usepackage{etoolbox}

\makeatletter
\def\@email#1#2{%
 \endgroup
 \patchcmd{\titleblock@produce}
  {\frontmatter@RRAPformat}
  {\frontmatter@RRAPformat{\produce@RRAP{*#1\href{mailto:#2}{#2}}}\frontmatter@RRAPformat}
  {}{}
}%
\makeatother
\begin{document}
\renewcommand{\topfraction}{0.95}
\renewcommand{\dbltopfraction}{0.95}
\renewcommand{\textfraction}{0.05}
\renewcommand{\floatpagefraction}{0.8}
\renewcommand{\dblfloatpagefraction}{0.8}
\setcounter{topnumber}{4}
\setcounter{dbltopnumber}{4}
\setcounter{totalnumber}{6}
\raggedbottom

\title{Nanowire-Assisted Epitaxial Lateral Overgrowth of Planar GaN(0001) on Diamond(100)}

\author{Brijmohan Paramanik*}
\email{brijmohan.paramanik@iitb.ac.in}
\affiliation{Department of Electrical Engineering, Indian Institute of Technology Bombay, Mumbai 400076, Maharashtra, India}
\author{Mayur Shinde}
\affiliation{Department of Electrical Engineering, Indian Institute of Technology Bombay, Mumbai 400076, Maharashtra, India}
\author{Subhranshu Sekhar Sahu}
\affiliation{Department of Electrical Engineering, Indian Institute of Technology Bombay, Mumbai 400076, Maharashtra, India}
\author{Soumyadip Chatterjee}
\affiliation{Department of Electrical Engineering, Indian Institute of Technology Bombay, Mumbai 400076, Maharashtra, India}
\author{Prajwal Pandey}
\affiliation{Department of Electrical Engineering, Indian Institute of Technology Bombay, Mumbai 400076, Maharashtra, India}
\author{Anshul Jain}
\affiliation{Center for Research in Nano Technology and Science, Indian Institute of Technology Bombay, Mumbai 400076, Maharashtra, India}
\author{Himanjal Bhattacharyya}
\affiliation{Department of Electrical Engineering, Indian Institute of Technology Bombay, Mumbai 400076, Maharashtra, India}
\author{Alpesh Donga}
\affiliation{VegaSemi Tech LLP, Surat , Gujarat, India}
\author{Apurba Laha}
\affiliation{Department of Electrical Engineering, Indian Institute of Technology Bombay, Mumbai 400076, Maharashtra, India}

\begin{abstract}
GaN-on-diamond heterostructures are promising for high-power and high-frequency electronics, however, direct epitaxial growth on diamond (100) remains challenging due to severe lattice and thermal mismatches, which induce high threading dislocation densities. To overcome this challenge, we have used Plasma Assisted Molecular Beam Epitaxy (PAMBE) to grow wurtzite GaN layers on diamond(100) substrates using nanowire-assisted epitaxial lateral overgrowth (ELOG). Two nanowire templates (NW$_{100}$, NW$_{600}$) were benchmarked against a nanowire-free reference (NW$_{000}$). High-resolution XRD, $\phi$-scans, (0002) pole figures, and reciprocal space mapping show that NW$_{100}$ attains the lowest dislocation density ($5.51 \times 10^{-4}$~nm$^{-2}$) and microstrain ($2.736 \times 10^{-3}$), the narrowest (0002) reflection, and a $c$-axis lattice parameter (5.1821~\AA) within 0.06\% of the bulk value; NW$_{000}$, in contrast, exhibits a contracted $c$-axis (5.1633~\AA) and elevated dislocation density ($6.26 \times 10^{-4}$~nm$^{-2}$), consistent with residual tensile stress. Cross-sectional TEM confirms the intended diamond/AlN/nanowire/ELOG/planar-GaN layer sequence and shows threading dislocations bending into near-lateral trajectories, directly evidencing defect filtering by the nanowire template. Together, these results establish nanowire-assisted ELOG as an effective strategy to suppress dislocation propagation and recover near-bulk lattice parameters in GaN-on-diamond heteroepitaxy, providing a structurally well-defined platform for device integration.
\end{abstract}

\maketitle

Gallium nitride (GaN) is a cornerstone semiconductor for high-power and high-frequency electronics, yet its device performance is constrained by self-heating under high-field operation. Integrating GaN with diamond, which combines exceptional thermal conductivity with mechanical robustness, has therefore emerged as a compelling route toward next-generation power and RF devices \cite{Zhang2025}. However, the attainable thermal benefit is governed largely by the interface, where interlayers and nucleation regions can introduce substantial thermal boundary resistance. \cite{Sun2015, Yates2018, Fan2025}

Direct epitaxial growth of GaN on diamond is thus highly desirable but remains challenging owing to the severe heteroepitaxial mismatch, difficult nucleation, polarity control, and defect formation. Continuous GaN films have been grown on single-crystal diamond (111) by plasma-assisted molecular beam epitaxy (PAMBE) using a thin AlN buffer and nitrogen-plasma pretreatment, which suppress inversion domains and enhance structural quality. \cite{Pecz2013} Growth on diamond (100) has also been demonstrated, although the emergence of multiple in-plane domains reflects the greater crystallographic complexity of this cubic orientation. \cite{vanDreumel2011}

Nanowire-mediated growth offers an attractive alternative to direct planar heteroepitaxy: the reduced interfacial footprint and free sidewalls of GaN nanowires promote strain relaxation and limit the propagation of extended defects. Self-assembled epitaxial GaN nanowires on diamond have been realized by PAMBE, \cite{Schuster2012} and selective-area growth on heteroepitaxial diamond (100) a cost-efficient and scalable substrate has yielded nanostructures oriented normal to the surface and exclusively Ga-polar, an advantage over growth on the (111) surface. \cite{Pantle2021} Building on these seeds, epitaxial lateral overgrowth (ELOG) provides not only planarization but also a transition from locally relaxed nanowires to a continuous upper GaN layer, mitigating the influence of the most defective region of the heterointerface. Although ELOG has demonstrated single-crystalline GaN growth on polycrystalline diamond, \cite{Ahmed2020} nanowire-assisted ELOG on single-crystal diamond (100) remains largely unexplored, leaving the effectiveness of nanowire templates in dislocation filtering and crystalline recovery unresolved.

In this Letter, we demonstrate wurtzite GaN films grown on diamond (100)(supplied by VegaSemi Tech LLP) by PAMBE using a nanowire-assisted ELOG approach, and quantify the resulting improvement in crystalline quality. This study shows that, relative to direct planar growth, the ELOG films exhibit reduced dislocation density and microstrain, a $c$-axis lattice parameter within 0.06\% of the bulk value, and lower surface roughness. These results establish nanowire-assisted ELOG as an effective route to filter threading dislocations and recover near-bulk lattice parameters on single-crystal diamond (100), providing a structurally well-defined platform for thermally managed GaN-on-diamond device integration.

\begin{figure}[!ht]
    \centering
    \includegraphics[width=\linewidth]{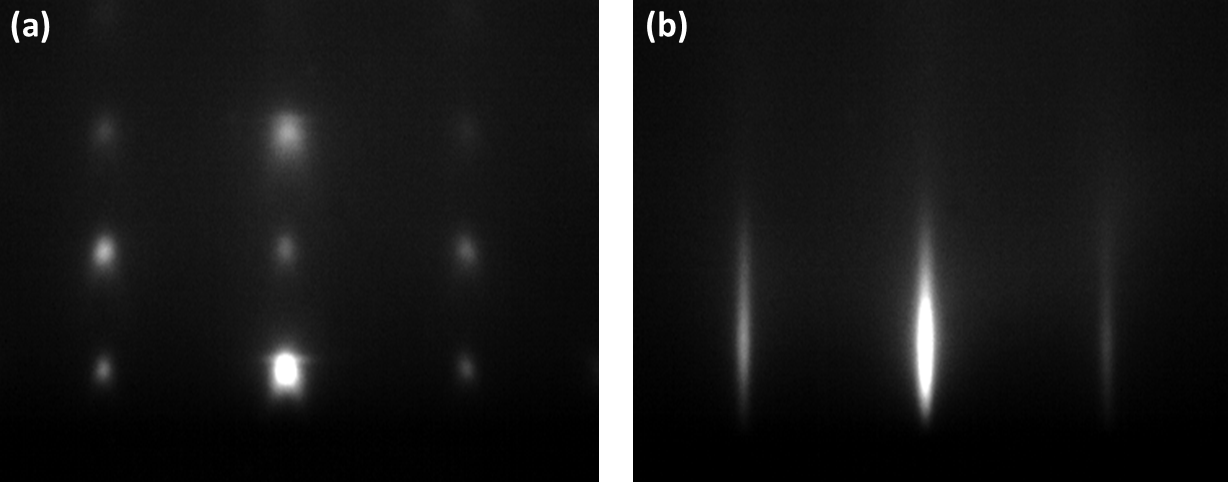}
    \caption{In-situ RHEED patterns recorded during growth, showing the transition from (a) spotty patterns characteristic of three-dimensional (nanowire) growth to (b) streaky patterns indicative of a smooth two-dimensional planar surface.}
    \label{fig:gaN_RHEED}
\end{figure}

The GaN heterostructures compared in this work were grown on single-crystalline 10$\times$10 mm$^2$ diamond (100) substrates in a Riber C21 molecular beam epitaxy system equipped with a Veeco UniBulb plasma source. Substrates were degassed at 820 {\textdegree}C, followed by deposition of a 20 nm AlN buffer at 800 {\textdegree}C and growth of GaN nanowires (NWs) at 780 {\textdegree}C. Three samples were prepared: NW$_{100}$ and NW$_{600}$ with 100 nm and 600 nm NWs, respectively, and a NW-free reference (NW$_{000}$) to assess the ELOG process. For NW$_{100}$ and NW$_{600}$, the NWs were subsequently subjected to epitaxial lateral overgrowth, in which the Ga/N ratio was increased and the growth temperature lowered to promote lateral expansion and coalescence into a planar layer. The transition from a spotty to a streaky RHEED pattern confirmed planar-layer formation (Fig. \ref{fig:gaN_RHEED}). Finally, 1 $\mu$m thick planar GaN layers were grown on all samples.

The crystalline structure, phase purity, and microstructure were characterized by high-resolution X-ray diffraction (HRXRD; Rigaku SmartLab, Cu K$\alpha$, $\lambda = 1.54059$~\AA, 40~kV, 30~mA) with a Ge $(220)$ two-bounce monochromator and a scintillation detector. From the (0002) reflection we extracted the microstrain (\( \varepsilon = \beta / 4 \tan \theta \)), $c$-axis lattice parameter (\( c = 2 d_{0002} \)), and dislocation density (\( \delta = 1/D^2 \)), where \( D \) follows from the Debye--Scherrer equation~\cite{Paramanik2022}.

In-plane orientation was obtained from $\phi$-scans of the asymmetric GaN $(\bar{1}01)$ reflection and out-of-plane texture from a $(0002)$ pole figure ($0^{\circ}$--$360^{\circ}$ azimuth, $0^{\circ}$--$80^{\circ}$ tilt, $1^{\circ}$ step). Reciprocal space maps (RSM) about the symmetric $(0002)$ reflection, resolved in $Q_\parallel$ and $Q_\perp$, provided the lateral registry, in-plane mosaicity, and strain state. All diffraction data were acquired at room temperature (Rigaku SmartLab Studio II). Cross-sectional imaging used an FEI Tecnai G2 F30 TEM at 300~kV, while surface morphology and roughness were assessed by JEOL JSM-7600F FEG-SEM and Oxford Instruments AFM, respectively.

\begin{figure*}[!ht]
    \centering
    \includegraphics[width=0.95\linewidth]{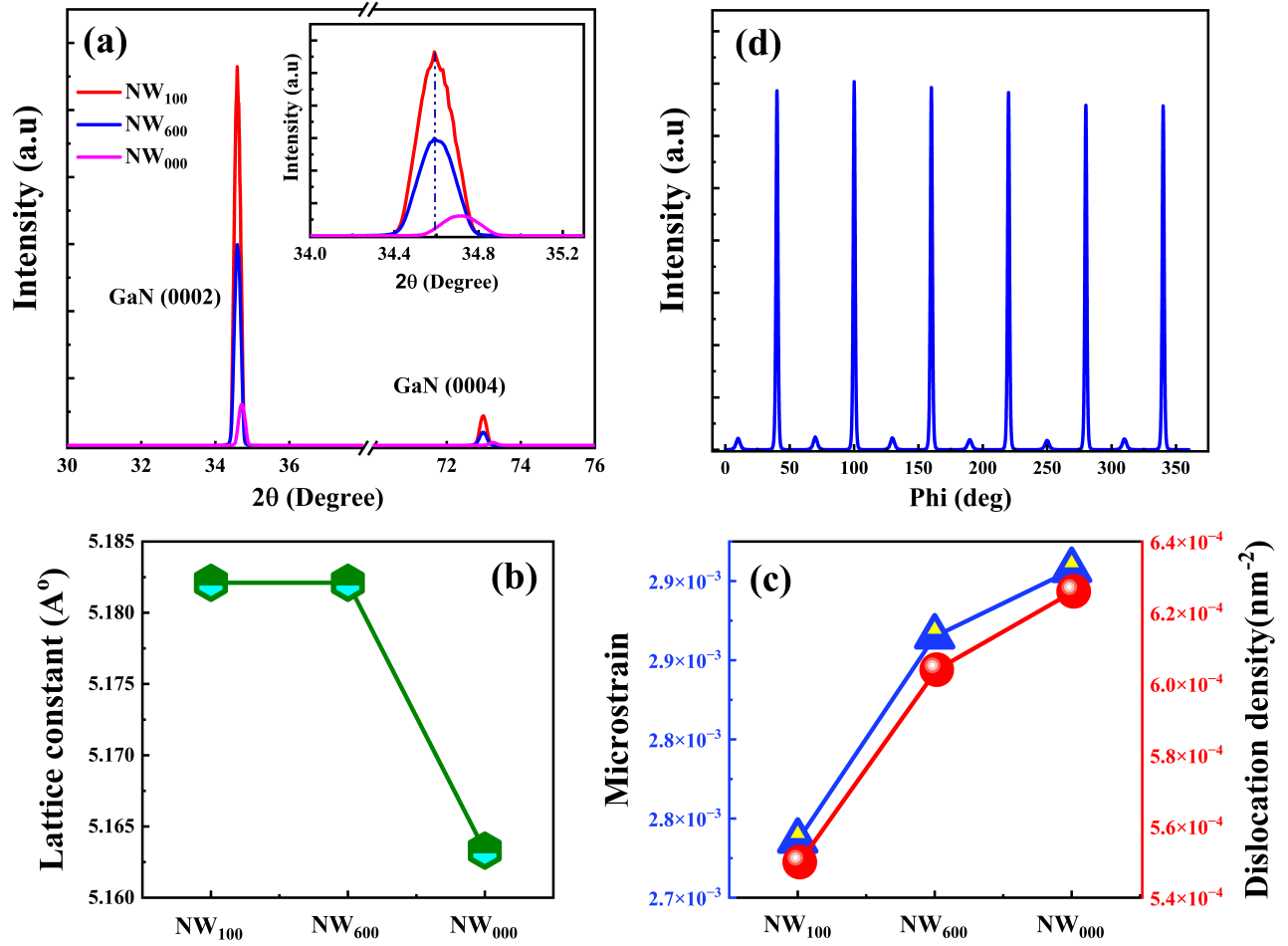}
    \caption{Structural characterization of GaN samples NW$_{100}$, NW$_{600}$, and NW$_{000}$: (a) XRD $\theta$--2$\theta$ patterns with the (0002) and (0004) wurtzite reflections (inset: enlarged (0002) peak); (b) lattice constant; (c) microstrain and dislocation density; and (d) azimuthal $\varphi$-scan of the (101) reflection for NW$_{100}$.}
    \label{fig:gaN_xrd}
\end{figure*}

Figure~\ref{fig:gaN_xrd}(a) shows the XRD $\theta$--$2\theta$ scans, where only the characteristic $(0002)$ and $(0004)$ reflections of hexagonal wurtzite GaN are observed, confirming a single $c$-axis out-of-plane orientation\cite{ReddyNallabala2024,Fakhri2023,Damasceno2025}. The (0002) intensity is highest for NW$_{100}$ and decreases in the order NW$_{100}$ $>$ NW$_{600}$ $>$ NW$_{000}$, a sequence reproduced by the (0004) reflection. The weak diffraction from NW$_{000}$ indicates a smaller volume fraction of coherently oriented GaN and/or a higher defect density, as expected for direct planar growth of GaN on diamond.

The inset resolves systematic differences in peak position and full width at half maximum (FWHM). The (0002) peaks of NW$_{100}$ and NW$_{600}$ lie close to the strain-free GaN position ($2\theta \approx 34.56^\circ$), whereas that of NW$_{000}$ is shifted to $\sim34.8^\circ$, indicating contraction of the out-of-plane lattice parameter $c$ and residual in-plane tensile strain \cite{Harutyunyan2001,Kim2025}. The correspondingly larger FWHM of NW$_{000}$ indicates greater diffraction broadening and higher microstrain, whereas the narrower lines of NW$_{100}$ and NW$_{600}$ indicate improved structural order.

The $c$-axis lattice parameter is 5.1821~\AA{} for both  NW$_{100}$ and  NW$_{600}$, differing by only ~0.06\% from the bulk wurtzite GaN value (5.185~\AA{}), whereas NW$_{000}$ yields 5.1633\,\AA{} [Fig.~\ref{fig:gaN_xrd}(b)]\cite{Yadav2022,Lee2009}. This out-of-plane contraction is consistent with the Poisson response to in-plane biaxial tensile stress and thus with a strained lattice state.

The microstrain increases from $2.736 \times 10^{-3}$ for NW$_{100}$ to $2.865 \times 10^{-3}$ for NW$_{600}$ and $2.907 \times 10^{-3}$ for NW$_{000}$ [Fig.~\ref{fig:gaN_xrd}(c)]. The concurrent increase in microstrain and contraction of $c$ in NW$_{000}$ indicate accumulation of structural defects, including threading dislocations, stacking faults, and point-defect complexes, generated during direct planar growth on diamond, where nucleation directly on the substrate limits island coalescence \cite{Darakchieva2007}. The dislocation density follows the same trend, rising from $5.51 \times 10^{-4}$\,nm$^{-2}$ for NW$_{100}$ to $6.04 \times 10^{-4}$\,nm$^{-2}$ for NW$_{600}$ and $6.26 \times 10^{-4}$\,nm$^{-2}$ for NW$_{000}$.

The three configurations therefore define a clear structural hierarchy. NW$_{000}$ (Planar GaN grown directly on diamond substrate) combines the highest dislocation density and microstrain with the broadest (0002) reflection: in the absence of a compliant nanowire template, the film accommodates the lattice and thermal mismatch with diamond through dislocation generation and residual elastic strain. NW$_{600}$ (600\,nm GaN nanowires/ELOG GaN/1\,$\mu$m planar GaN) is intermediate; sidewall relaxation of the taller nanowires provides partial dislocation filtering, but its higher microstrain and larger (0002) linewidth relative to NW$_{100}$ indicate less efficient defect annihilation during ELOG coalescence. However, NW$_{100}$ (100\,nm GaN nanowires/ELOG GaN/1\,$\mu$m planar GaN) exhibits the narrowest (0002) reflection, the lowest microstrain and dislocation density, and the $c$-axis parameter closest to strain-free bulk GaN, consistent with more effective lateral strain relaxation at the thinner template and additional coalescence-induced dislocation annihilation during ELOG.

Because these metrics describe only the out-of-plane response, the in-plane epitaxial relationship was examined. The $\phi$-scan of the asymmetric (101) reflection of NW$_{100}$ [Fig.~\ref{fig:gaN_xrd}(d)] shows six maxima spaced by $60^\circ$, reproducing the hexagonal rotational symmetry of wurtzite GaN and confirming in-plane epitaxial alignment. The absence of additional or split azimuthal features indicates that rotational domains and incoherently oriented regions are suppressed, and that the nanowire/ELOG layer is in-plane coherent with the diamond/AlN stack. The narrow and nearly equal-intensity poles further indicate a low in-plane mosaic spread and a laterally homogeneous crystallographic orientation across the coalesced film \cite{Kushvaha2013, Shetty2013}.

\begin{figure}[htbp]
    \centering
    \includegraphics[width=0.7\linewidth]{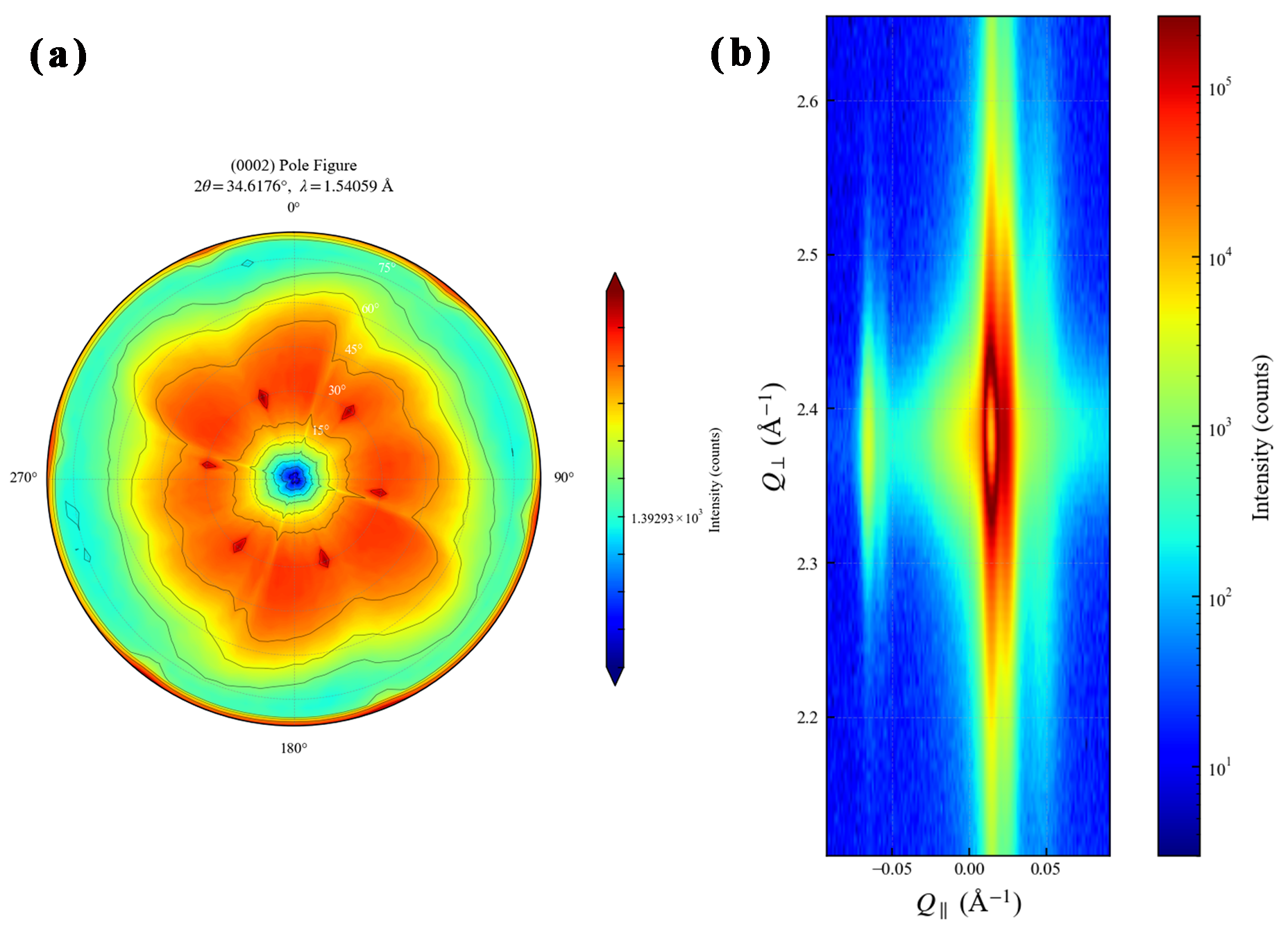}
    \caption{out-of-plane texture of NW$_{100}$: (a) (0002) pole figure showing $c$-axis fiber texture; (b) (0002) RSM showing a single, $Q_{\perp}$-elongated reciprocal lattice point.}
    \label{fig:pole_figure}
\end{figure}

The (0002) pole figure of NW$_{100}$ [Fig.~\ref{fig:pole_figure}(a)] shows an intensity maximum confined to the center of the stereographic projection that decays steeply with tilt angle, indicating a pronounced $c$-axis fiber texture in the GaN layer grown on diamond through the AlN nucleation layer and the nanowire/ELOG buffer. The limited angular spread of the central pole corresponds to small tilt deviations of the GaN $c$ axes from the surface normal, in agreement with the narrow rocking-curve widths measured across the 1\,$\mu$m planar GaN layer \cite{Liu1998,Freitas2025}. The nearly circular azimuthal distribution, weakly modulated by the sixfold in-plane symmetry, indicates a single predominant in-plane orientation with minimal residual mosaicity despite the large lattice and thermal mismatch between GaN and diamond~\cite{Schulze2004}, confirming that the nanowire template combined with lateral overgrowth mitigates the heteroepitaxial mismatch and limits threading-dislocation propagation.

The structural coherence of the GaN layer in NW$_{100}$ was examined further by high-resolution reciprocal space mapping (RSM) around the symmetric (0002) reflection, plotted in the reciprocal coordinates $Q_{\parallel}$ and $Q_{\perp}$ [Fig.~\ref{fig:pole_figure}(b)]. A single reciprocal lattice point centered at $Q_{\parallel} \approx 0$ indicates lateral registry between the GaN epilayer and the underlying AlN/diamond stack, with no resolvable macroscopic tilt or in-plane relaxation \cite{Paskova2004}. The intensity is elongated along $Q_{\perp}$ and narrow along $Q_{\parallel}$, consistent with a $c$-axis-oriented epilayer of low mosaic tilt and small in-plane lattice-parameter dispersion; the out-of-plane broadening is attributed to the finite film thickness and to a spread in lattice spacing associated with residual strain and threading dislocations. No secondary maxima or shoulders appear along $Q_{\parallel}$, indicating that the GaN overgrowth behaves as a single epitaxial domain rather than a mixture of relaxed and strained regions. Together with the $\phi$-scan and pole-figure, the RSM establishes that nanowire-templated ELOG of GaN on diamond yields a structurally coherent, nearly strain‑free epilayer with minimized in‑plane mosaic spread and well‑defined out‑of‑plane lattice parameters, providing a viable platform for GaN-on-diamond device integration~\cite{Pereira2002}.

\begin{figure*}[!ht]
\centering
\includegraphics[width=\textwidth]{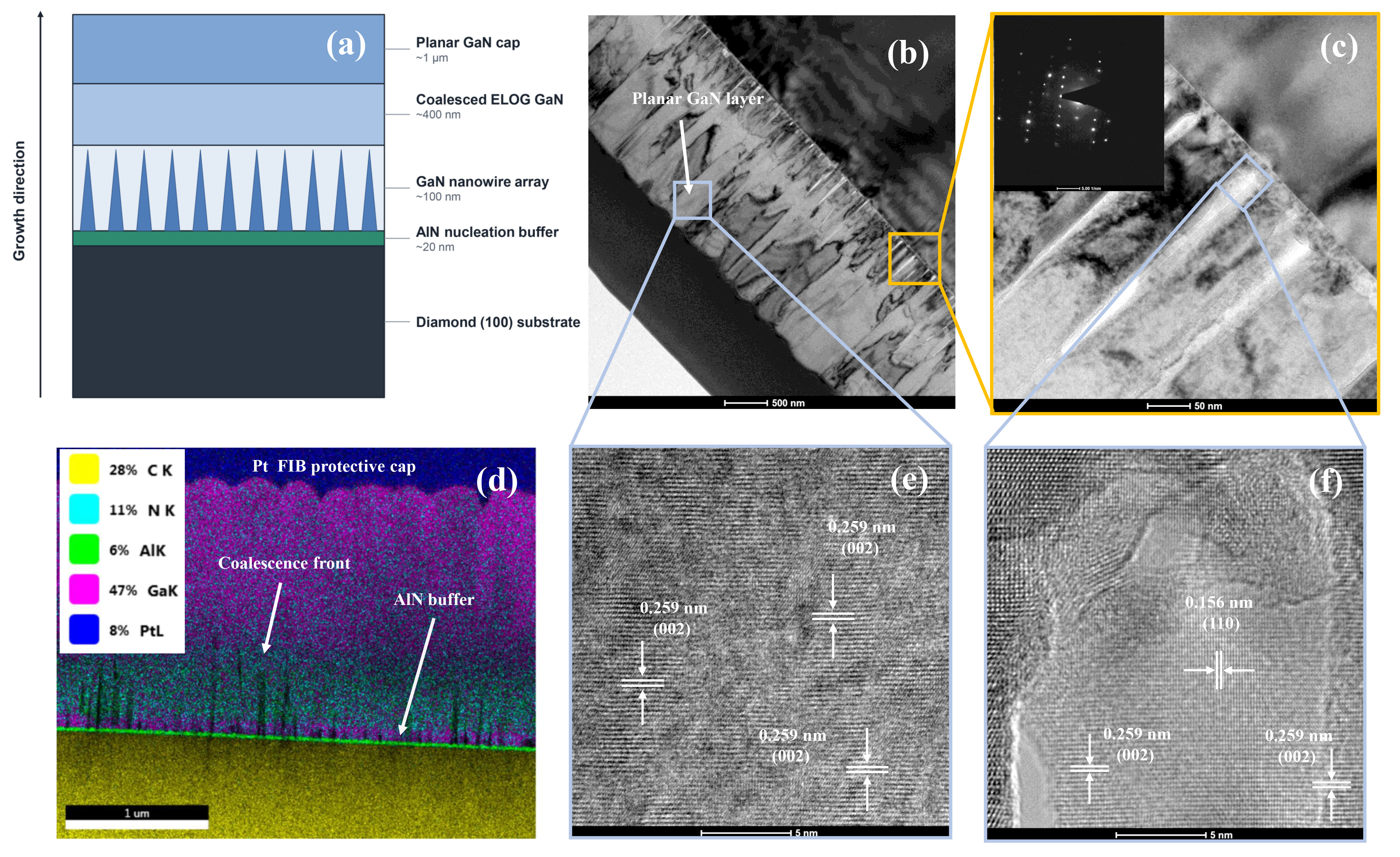}
\caption{Cross-sectional NW\textsubscript{100} microstructure: (a) Schematic diagram, (b) Bright-field TEM of the stack, (c) High-magnification TEM of the GaN nanowire layer (inset: SAED). (d) EDX mapping (e) HRTEM of the upper planar region, and (f) HRTEM of a Nanowire base.}
\label{TEM}
\end{figure*}

While diffraction establishes high crystalline quality, direct imaging is required to identify the regions where dislocation filtering improves lattice integrity. Cross-sectional TEM confirms the intended layer sequence and reveals the microstructural origin of the low defect density in NW\textsubscript{100}. Figure~\ref{TEM}(a) shows the architecture schematically, and the low-magnification bright-field image in Fig.~\ref{TEM}(b) resolves the complete stack along the growth direction. The diamond (100) substrate appears uniformly dark and featureless, consistent with a thick, defect-free single crystal, while the \SI{20}{nm} AlN buffer, near the contrast limit at this magnification, is resolved unambiguously in the elemental map of Fig.~\ref{TEM}(d), confirming the diamond/AlN interface. Within the continuous GaN epilayer, diffraction contrast separates the buried template from the planar cap: the lower region, comprising the \SI{100}{nm} nanowire array and the \SI{400}{nm} coalesced ELOG layer, concentrates the near-vertical extended-defect contrast originating at the heterointerface, along with curved bend contours from lamella flexure during FIB thinning. In contrast, the upper \SI{1}{\micro\meter} planar GaN layer exhibits uniform, laterally continuous contrast up to the free surface with no evident propagation of the underlying defects, providing direct evidence of defect filtering by the nanowire-assisted ELOG template.

Figure~\ref{TEM}(c) is a higher-magnification image of an individual GaN nanowire from the seed array outlined in Fig.~\ref{TEM}(b), resolving its columnar morphology and straight sidewalls. The corresponding SAED pattern (inset) shows only discrete reflections of a single reciprocal lattice, with no Debye--Scherrer rings or rotated spot families, confirming single-crystalline wurtzite GaN nucleation without rotational domains~\cite{Muhammed2016}. The composite EDX map in Fig.~\ref{TEM}(d) independently verifies the chemical sequence: C is confined to the substrate, Al forms a narrow continuous interfacial band indicating an unbroken AlN buffer with no direct GaN--diamond contact, Ga fills the overlayer, and Pt is restricted to the FIB protective cap. The Ga distribution further reproduces the microstructure, with Ga-deficient striations terminating at a laterally continuous coalescence front beneath a uniform signal, indicating that residual inter-nanowire voids remain confined to the buried template while the planar GaN layer is fully dense.

The defect-filtering mechanism underlying the low dislocation density of NW\textsubscript{100} is evident in the contrast evolution of Fig.~\ref{TEM}(b): threading dislocations nucleated at the heterointerface bend out of the growth direction into steeply inclined, near-lateral trajectories. Such bending promotes termination at free surfaces and mutual annihilation of segments with opposite Burgers vectors, lowering the dislocation flux into the upper epilayer without altering the interfacial dislocation density. HRTEM resolves two structurally distinct regions of the film. At the nanowire base [Fig.~\ref{TEM}(f)], continuous lattice fringes span the highly mismatched interface with no detectable amorphous layer: one domain resolves \SI{0.259}{nm} (0002) fringes, while an adjacent domain shows \SI{0.156}{nm} $\{11\bar{2}0\}$ fringes at a large angle to (0002), consistent with imaging along the $\langle 1\bar{1}00\rangle$ zone axis. The upper planar region [Fig.~\ref{TEM}(e)] comprises three adjacent domains sharing a common \SI{0.259}{nm} (0002) spacing with close angular orientation, corresponding to $c = \SI{0.518}{nm}$, in agreement with the HRXRD value of \SI{5.1821}{\angstrom}. A slight residual misorientation is thus accommodated by low-angle tilt boundaries rather than by a distributed threading-dislocation network. This evolution from a fully crystalline nanowire base to a uniformly (0002)-oriented planar layer reflects coalescence-driven structural ordering during growth and accounts for the low microstrain ($2.736 \times 10^{-3}$ ) and low dislocation density ($5.51 \times 10^{-4}$\,nm$^{-2}$) of NW\textsubscript{100}.

\begin{figure}[!ht]
    \centering
    \includegraphics[width=\columnwidth]{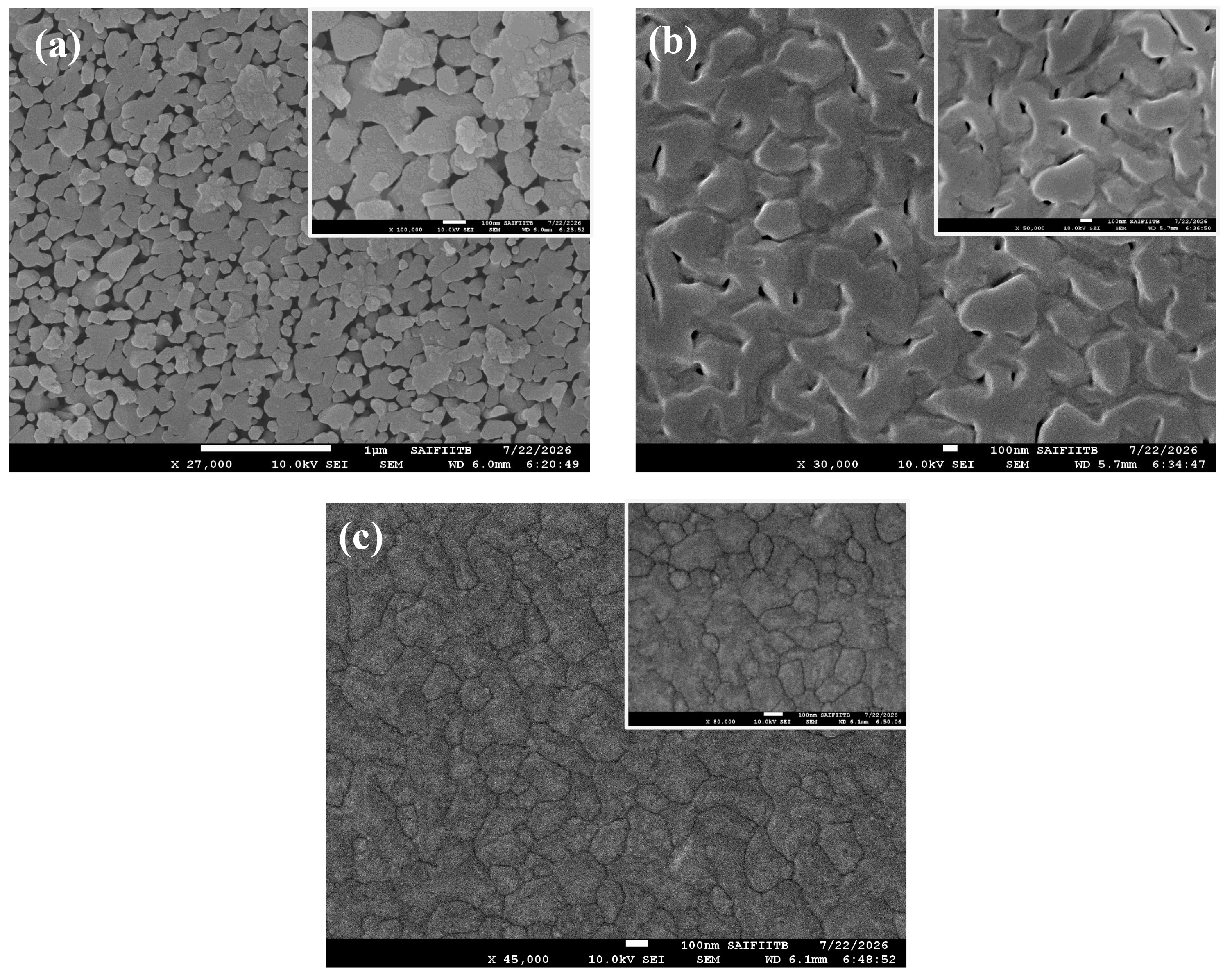}
    \caption{Plan-view SEM micrographs of (a) NW$_{100}$, (b) NW$_{600}$, and (c) NW$_{000}$; insets show the corresponding higher-magnification views.}
    \label{fig:sem_micrographs}
\end{figure}

These buried microstructural differences are also studied at the free surface, as demonstrated by FE-SEM [Fig.~\ref{fig:sem_micrographs}]. For NW$_{100}$ [Fig.~\ref{fig:sem_micrographs}(a)], the planar GaN cap over the 100~nm nanowire/ELOG template forms a continuous film of densely packed, laterally merged domains with straight-edged polyhedral terminations (inset); the absence of open porosity or V-shaped pits indicates complete coalescence above the 20~nm AlN nucleation layer~\cite{Jung2008}. For NW$_{600}$ [Fig.~\ref{fig:sem_micrographs}(b)], the same cap over a 600~nm template yields a coalesced matrix decorated with dark, angular pits with V-shaped sidewalls (inset) signify open-core threading dislocations and unclosed coalescence boundaries; their elongation along a common in-plane direction indicates that the taller template exceeds the lateral mass-transport length during ELOG, so that adjacent growth fronts converge with residual angular mismatch. In contrast, NW$_{000}$ [Fig.~\ref{fig:sem_micrographs}(c)], grown directly on diamond without AlN or a nanowire/ELOG intermediate, exhibits closely packed sub-100~nm islands separated by a dense boundary network, with no extended coalescence domains or facets. This mosaic reflects Volmer--Weber-like growth driven by the severe lattice and thermal mismatch, in which each island boundary persists as a structural discontinuity through the film thickness.

\begin{figure}[htbp]
    \centering
    \includegraphics[width=1\linewidth]{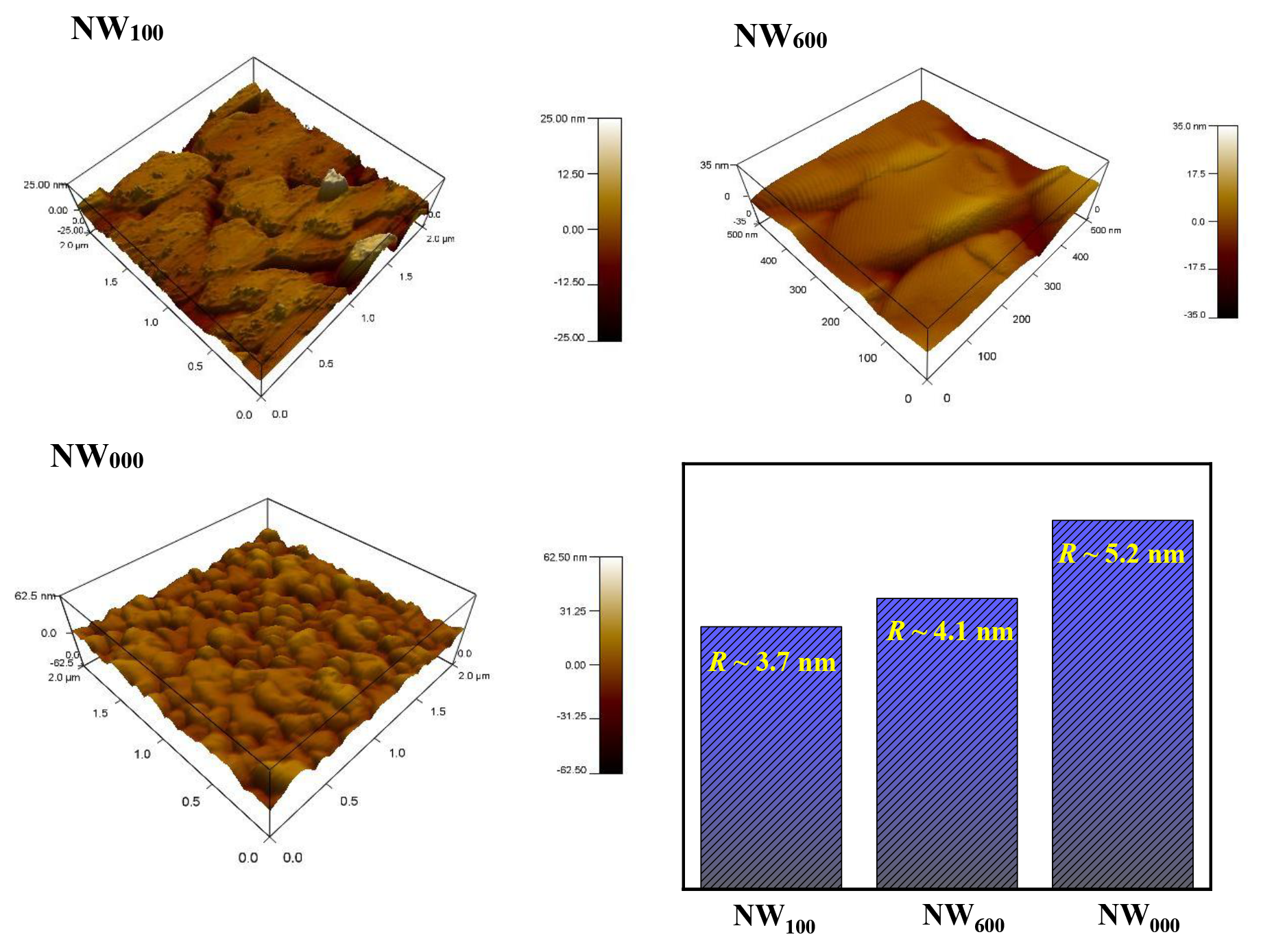}
    \caption{Three-dimensional AFM surface morphologies of samples NW$_{100}$, NW$_{600}$, and NW$_{000}$, with a bar graph showing the corresponding RMS roughness values.}
    \label{AFM}
\end{figure}

To quantify these morphological differences, the surface topography of the GaN/diamond heterostructures was examined by tapping-mode AFM; the three-dimensional height profiles and RMS roughness values are shown in Fig.~\ref{AFM}. For NW$_{100}$, the GaN overgrowth forms wide, laterally merged terraces bounded by shallow step edges with only a few localized asperities, giving the lowest RMS roughness, $R \sim 3.7$\,nm. This terrace-like morphology is consistent with step-flow or quasi-two-dimensional growth, a regime correlated with reduced threading-dislocation density and improved structural and optical quality \cite{QhalidFareed2000}.

The NW$_{600}$ surface shows elongated ridges aligned with the ELOG propagation direction and a higher RMS roughness of about 4.1\,nm, indicating more pronounced step bunching and less complete planarization of the lateral overgrowth front at this nanowire height \cite{Hino2000}. NW$_{000}$, which lacks the nanowire/ELOG layers, displays a closely packed island morphology with large height modulation and the highest RMS roughness ($\sim5.2$\,nm), characteristic of island-dominated growth with limited lateral coalescence. Therefore, the AFM results are consistent with the XRD structural metrics, identifying the 100~nm GaN nanowire template prior to ELOG as the optimal configuration for achieving the smoothest coalesced GaN surface.

\FloatBarrier
In summary, wurtzite GaN epilayers were grown on diamond (100) by PAMBE using \SI{100}{nm} and \SI{600}{nm} GaN nanowire/ELOG templates (NW$_{100}$ and NW$_{600}$) and a nanowire-free reference (NW$_{000}$). All samples were phase-pure wurtzite GaN with strong $c$-axis orientation, confirmed by exclusive (000$l$) $\theta$--$2\theta$ reflections. NW$_{100}$ exhibited the most favorable microstructure: the narrowest (0002) reflection, the lowest microstrain ($2.736 \times 10^{-3}$) and dislocation density ($5.51 \times 10^{-4}$~nm$^{-2}$), and a $c$-axis parameter (\SI{5.1821}{\angstrom}) within 0.06\% of the bulk value. NW$_{000}$, which lacks a defect-filtering interlayer, instead showed a contracted $c$ axis (\SI{5.1633}{\angstrom}) with elevated microstrain and dislocation density, consistent with unconstrained threading-dislocation propagation and residual tensile biaxial stress.

$\phi$-scans of NW$_{100}$ resolved six $60^\circ$-spaced poles with narrow linewidths, and pole-figure and RSM analyses yielded a single reciprocal lattice point at $Q_{\parallel} \approx 0$, indicating lateral coherence and effective strain accommodation by the nanowire/ELOG architecture. Cross-sectional and high-resolution TEM verified the intended layer sequence and an unbroken AlN buffer, and resolved threading dislocations bending into near-lateral trajectories that terminate at free surfaces or voids, thereby reducing the dislocation flux into the planar cap. Plan-view SEM distinguished three growth regimes complete coalescence for NW$_{100}$, residual V-pits for NW$_{600}$, and Volmer--Weber island growth for NW$_{000}$ consistent with the RMS roughness measured by AFM, which increased from \SI{3.7}{nm} for NW$_{100}$ to \SI{5.2}{nm} for NW$_{000}$.

These results identify the \SI{100}{nm} GaN nanowire/ELOG template as the configuration that most effectively accommodates the lattice and thermal-expansion mismatch in GaN-on-diamond heteroepitaxy, simultaneously reducing dislocation density and in-plane mosaicity while yielding the smoothest coalesced surface. The NW$_{100}$ architecture thus provides a structurally well-defined platform for GaN-on-diamond high-power and high-frequency device integration.

\par
The authors acknowledge financial support from the Ministry of Electronics and Information Technology (MeitY), Government of India, through the TaRANG Project (Project ID: TaRANG), and the IIT Bombay Nanofabrication Facility.
\section*{AUTHOR DECLARATIONS}

\subsection*{Conflict of Interest}
The authors have no conflicts to disclose.

\subsection*{Data Availability}
Data will be made available on request.
\vspace{0.5cm}

\textbf{REFERENCES}
\bibliography{references} 
\end{document}